\documentclass{aastex631}

\usepackage[T1]{fontenc}
\usepackage{graphicx}	% Including figure files
\usepackage{amsmath}	% Advanced maths commands
\usepackage{amssymb}	% Extra maths symbols
\usepackage{hyperref}
\usepackage{makecell}
\usepackage{color} 		% Colors for text commenting
\usepackage{comment}    % Allows for commenting out blocks of text

\newcommand{\nuc}{h\nu_{\rm c}}
\newcommand{\num}{h\nu_{\rm m}}
\newcommand{\gc}{\gamma_{\rm c}}
\newcommand{\gm}{\gamma_{\rm m}}
\newcommand{\gb}{\Gamma_{\rm boost}}
\newcommand{\grel}{\gamma_{\rm rel}}
\newcommand{\me}{m_{\rm e}}
\newcommand{\mpp}{m_{\rm p}}
\newcommand{\F}{F_{\nu, \rm c}}
\newcommand{\Pn}{P_{\nu, {\rm p}}}
\newcommand{\simleq}{\; \raisebox{-0.4ex}{\small$\stackrel
{{\textstyle<}}{\sim}$}\;}

\definecolor{BA}{rgb}{1,0.6,0 }

\newcommand{\R} {\text{RMS}}

\begin{document}

\label{firstpage}

\title{Onset of particle acceleration during the prompt phase in gamma-ray bursts as revealed by  synchrotron emission in GRB160821A
}

\correspondingauthor{Shabnam Iyyani}
\email{shabnam@iisertvm.ac.in}
\correspondingauthor{Felix Ryde}
\email{fryde@kth.se}

\author[0000-0002-9769-8016]{Felix Ryde}
\affiliation{Department of Physics, KTH Royal Institute of Technology, \\
and The Oskar Klein Centre, SE-10691 Stockholm, Sweden}

\author{Shabnam Iyyani}
\affiliation{Indian Institute of Science Education and Research, \\
Thiruvananthapuram, 695551, Kerala, India}

\author{Bj\"orn Ahlgren}
\affiliation{Department of Physics, KTH Royal Institute of Technology, \\
and The Oskar Klein Centre, SE-10691 Stockholm, Sweden}

\author{Asaf Pe'er}
\affiliation{Department of Physics, Bar-Ilan University, Ramat-Gan 52900, Israel}

\author{Vidushi Sharma}
\affiliation{Department of Physics, KTH Royal Institute of Technology, \\
and The Oskar Klein Centre, SE-10691 Stockholm, Sweden}

\author[0000-0002-0642-1055]{Christoffer Lundman}
\affiliation{Department of Physics, KTH Royal Institute of Technology, \\
and The Oskar Klein Centre, SE-10691 Stockholm, Sweden}

\author[0000-0002-9769-8016]{Magnus Axelsson}
\affiliation{Department of Physics, KTH Royal Institute of Technology and The Oskar Klein Centre, 10691 Stockholm, Sweden}

\begin{abstract}
% {Gamma-ray bursts (GRBs) are among the most powerful explosions in the Universe and predominantly produce gamma-rays for a short duration, ranging from a fraction of a second to several 100’s of seconds. 
{
{The physical processes of the gamma-ray emission and particle acceleration during the prompt phase in gamma ray bursts (GRBs) are still unsettled. In order to perform an unambiguous physical modelling of observations, a clear identification of the emission mechanism is needed.}
An instance of a clear identification is the synchrotron emission during the very strong flare in GRB160821A, that  occurs during the prompt phase at 135\,s. Here we show that the distribution of the radiating electrons in this flare is initially very narrow, but later develops a power-law tail of accelerated electrons. We thus identify for the first time the onset of particle acceleration in a GRB jet. The flare is consistent with a late energy release from the central engine causing an external-shock as it encounters a preexisting ring nebula of a progenitor Wolf-Rayet star.  
Relativistic forward and reverse shocks develop, leading to two distinct emission zones with similar properties. 
The particle acceleration only occurs in the forward shock, moving into the dense nebula matter. {Here, the magnetisation also decreases below the critical value, which allows for Fermi acceleration to operate. Using this fact,  we find a bulk Lorentz factor of $420  \simleq \Gamma  \simleq 770$, and an emission radius of $R \sim 10^{18}$ cm, indicating a tenuous gas of the immediate circumburst surrounding.} The observation of the onset of particle acceleration thus gives new and independent constraints on the properties of the flow as well as on theories of particle acceleration in collisionless astrophysical shocks.
%%%%%%%%
%%%%%%%%%
%The decrease in the magnetisation is found to be correlated with the onset of particle acceleration. 
}
\end{abstract}

\section{Introduction}

Particle acceleration is expected to occur in the relativistic, collisionless shocks in gamma-ray bursts \citep{Rees_Meszaros1994, Spitkovsky2008}. However, many aspects of the acceleration mechanism are not yet fully understood. Such aspects include the microphysical processes that mediate the acceleration of electrons, the physical conditions for such a process to become efficient, and the fraction of electrons that undergo acceleration.  While synchrotron emission from the external shock during the GRB afterglow reveals the power-law distribution of shock-accelerated particles \citep{Tavani1996, WijersGalama1999}, the situation for the prompt phases is less certain. Most prompt spectra have an exponential cut-off above their peak \citep{Goldstein2012, Yu2019}, which indicates that any particle acceleration is inefficient. On the other hand, some spectra have prominent high-energy power-law spectra above their peak \citep[e.g.][]{080916C, Axelsson2012}. 
At the same time, much evidence points towards that both synchrotron and emission from the jet photosphere contribute to a varying degree during the first few 100 seconds of a GRB emission \citep{Meszaros_etal_2002, Ajello2019, Li2020}. %(M\'esz\'aros \& Rees 2002, Ajello et al. 2019, Li et al. 2021).
In contrast to the synchrotron spectrum, photospheric emission spectra probe radiation mediated shocks \citep{Beloborodov2017, Samulesson2022} and therefore are related to a different physical setting.   Thus, correctly identifying the emission mechanism as being synchrotron is necessary to be able to identify and study any particle acceleration. % study the properties of the GRB jet. 

During the intense burst GRB~160821A \citep{Sharma_etal_2019}, synchrotron emission
is clearly identified since it has a broad, non-thermal spectrum with several breaks, at around 100 keV, 1000 keV, and 50 000 keV, which characterises synchrotron spectra of other GRBs \citep{Ogan2017, Acuner2018}.
The main emission also occurs later than 100\,s after the trigger and has a long duration which supports a synchrotron interpretation \citep{Ogan2019, Li2020}.
Other facts in support of synchrotron emission are its high degree of polarisation ($\ge 60\%$, in the energy range 100 - 300 keV) \citep{Sharma_etal_2019, Gill_Granot_Kumar_2020} and that it is very bright \citep{Ogan2017, Acuner2018}. In this paper, we therefore use synchrotron spectral fits of the  prompt emission in GRB160821A to study  distribution of the radiating electrons. 

The observed synchrotron emission is powered by energy dissipation in shocks, where the electrons are heated and cool rapidly in a local magnetic field $B$ \citep{Rees_Meszaros1994}. 
The electrons assume a quasi-Maxwellian energy distribution around a Lorentz factor $\gamma_{\rm m}$ corresponding to some fraction of the available dissipated energy. If the conditions are right, the electrons can be further accelerated in the shocks \citep{Sironi2011} forming a power-law distribution with an index $p\sim 2.2$ {\textendash} $2.5$, extending to higher energies ($N_{\rm el} (\gamma)\, d\gamma \propto \gamma ^{-p}$, above $\gamma_{\rm m}$).  Since the episode analysed here is very bright the emission has to be very efficient, which corresponds to that the cooling time of the radiating electrons has to be shorter than the typical dynamical time. This will cause a  distribution of cooled electrons $N_{\rm el} (\gamma)\, d\gamma \propto \gamma ^{-2}$ below $\gamma_{\rm m}$ down to a Lorentz factor of $\gamma_{\rm c}$, which depends on the magnetic field strength. Moreover, any high-energy power-law of accelerated electrons will become steeper by unity, to an index  of $p+1$ \citep[e.g., ][]{Sari_etal_1998}. 
As the electrons radiate the observed synchrotron photon spectrum will have corresponding power-law segments with breaks at energies $\nuc$ and $\num$.

\section{Synchrotron spectral fits of the Strong flare in GRB160821A}

 GRB160821A was observed %on 21 August 2016 
by several space observatories, among others 
%Swift BAT (Siegel et al. 2016), 
{\it AstroSat} \citep{160821A_astrosat} and the {\it Fermi gamma-ray space telescope} \citep{160821A_Fermi} (hereafter, {\it Fermi}).
%, Konus-Wind (Kozlova et al. 2016) and Gamma ray burst monitor on board CALET (Marrocchesi et al. 2016). 
 It is the third brightest GRB observed by {\it Fermi} in terms of energy flux observed in the energy range 10 - 1000 keV.  
%(see
%In Sharm 
The observed prompt emission of GRB160821A consists of two emission episodes where the first emission episode extends for a period of 112\,s from the time of trigger, and the second emission episode peaks at around 135\,s, lasting for around 40 s, and is nearly hundred times  brighter than the first emission episode (Figure 1). Here, we focus the study on this intense flare, i.e, the second episode. For the spectral analysis we choose the data ranging between roughly $8 \, \rm keV$ and $ 40 \, \rm MeV$ from {\it Fermi} Gamma-ray Burst Monitor (GBM) including sodium iodide (NaI)  and bismuth germanate (BGO) detectors (NaI6, NaI7, NaI9, BGO 1; \citet{Meegan2009_GBM}). In addition, the Large Area Telescope Low-Energy (LAT-LLE) and LAT data in the energy ranges $30 \, \rm MeV -130\, \rm MeV$ and $100 \, \rm MeV - 5 \, \rm GeV$ respectively are also used for the spectral analysis \citep{Atwood2009}. The same spectral files generated for the time resolved spectroscopy in \citet{Sharma_etal_2019} are used for this study. The effective area correction factors estimated in that study for the different detectors with respect to BGO 1 whose value was fixed to unity are the following: $0.97 \pm 0.01$ for n6, $0.92 \pm 0.01$ for n7, $0.94 \pm 0.01$ for n9 and $0.84 \pm 0.06$ for LAT. The spectral analysis is carried out in the Multi-Mission Maximum Likelihood (3ML) software \citep{Vianello2015}, wherein the synchrotron emission model \citep{Aharonian2010} is implemented using the NAIMA package \citep{Naima2015}.

We  divide the light curve of the main episode into three time bins, shown in the uppermost panel of Fig. 1. This division follows the one made for the polarisation measurement of \citet{Sharma_etal_2019}. They further showed that the spectral shapes are different, but relatively steady within these three intervals. This fact  further motivates to use the integrated signal during them.
Each time bin is fitted with a synchrotron spectrum, using a Bayesian analysis, with priors on the free parameters as described in Appendix \ref{app:1}. The right-hand  panel in Figure \ref{fig:1} shows the  best fit power spectrum ($\nu F_{\nu}$) for the three time intervals.  In each interval we thus determine the synchrotron cooling frequency, $\nuc$, the synchrotron frequency of the injected electrons, $\num$. In interval 2 we also identify a high-energy cutoff at $h \nu_{\rm cutoff}$ and the  high-energy powerlaw index. The parameter values are given in Table~\ref{tab:1}. The left-hand panel in Figure \ref{fig:1}  shows the corresponding energy  distribution of the radiating electrons. In appendix \ref{app:1} we further show the fitted spectra in count space (Fig. \ref{fig:fits_wMAP}), as well as the corner plot of all the fitted parameters, $\gm$, $norm$, and $p$ (Fig. \ref{fig:cornerplot_large}). %

We identify a few important spectral changes between the fits of the three timebins. While the first and third intervals are rather similar, the spectral shape of the second interval differs significantly. First, the ratio  $\gm/\gc$ is much smaller, second, a clear power-law distribution above $\gm$ is formed with $p=2.3$, and third, the flux level is the largest. It is interesting to note that the variations detected in interval 2 coincide with the change in polarisation degree \citep{Sharma_etal_2019}. 
The main conclusion from this spectral analysis is therefore that something happens in Interval 2 that is responsible for the onset of particle acceleration. 

\begin{figure*}
\begin{center}
\includegraphics[width=\columnwidth]{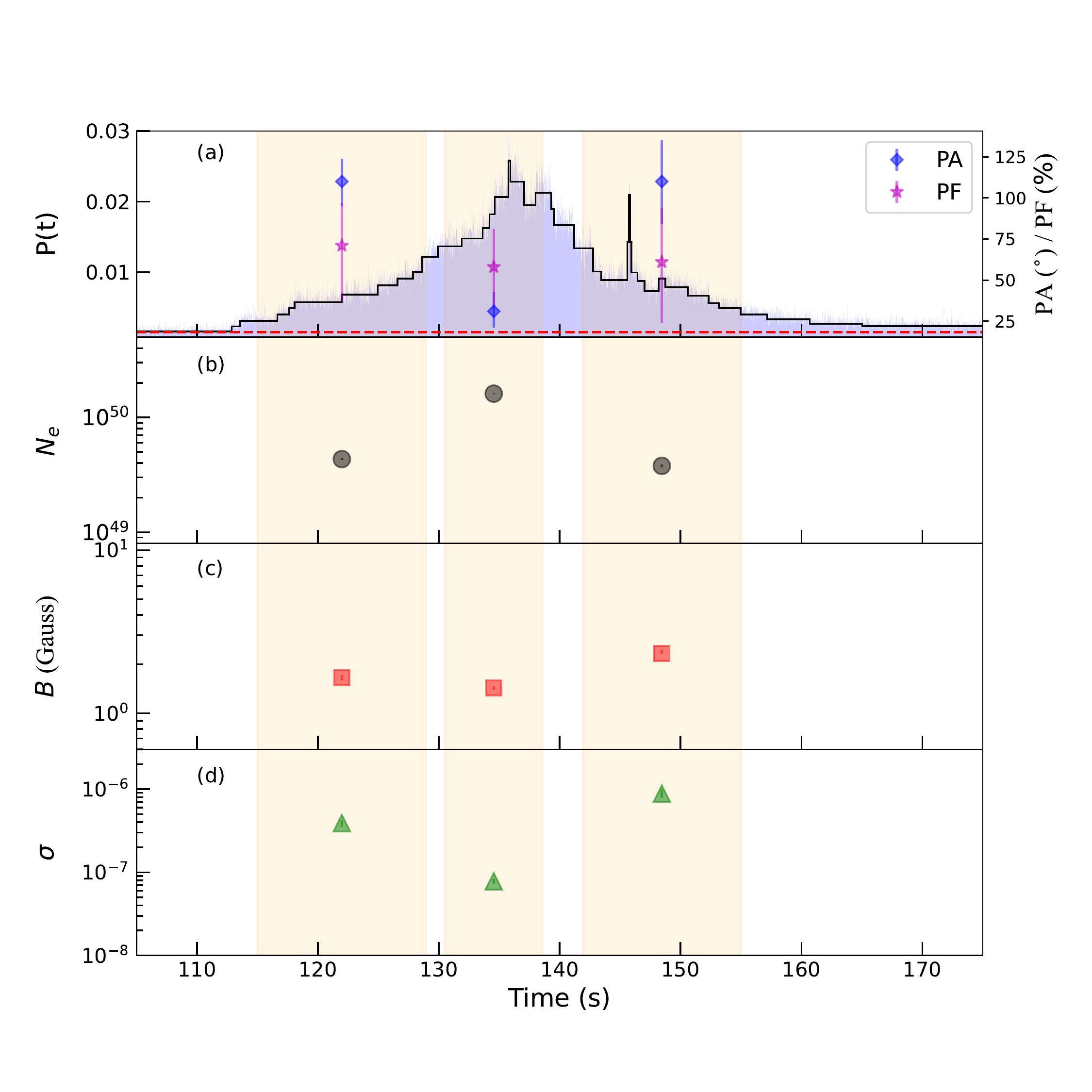}
\caption{{Evolution of the main episode of GRB~160821A.} The yellow shaded regions represent the three timebins that are analysed and correspond to time intervals where the polarisation measurements are made using {\it AstroSat} CZTI data. {\bf (a)} The high resolution ($0.01$ s) light curve (blue) and the Bayesian block binned light curve (black line), with a false alarm probability to compute the prior, $p_{0} = 0.01$. The y-axis represents the probability density which gives the counts per bin divided by the width of the bin. The polarisation fraction, PF, and the polarisation angle, PA, obtained \citet{Sharma_etal_2019} are shown in magenta star and blue diamonds respectively. The temporal variation of the derived physical parameters, assuming $\Gamma=300$ and $z=0.4$ {\bf (b)} the number of radiating electrons, $N_{\rm e}$; {\bf (c)} the co-moving magnetic field, $B$, and {\bf (d)} the magnetisation, $\sigma$, are shown. }
\label{fig:lc}
\end{center}
\end{figure*}

\begin{figure*}
\begin{centering}
\includegraphics[width=\columnwidth]{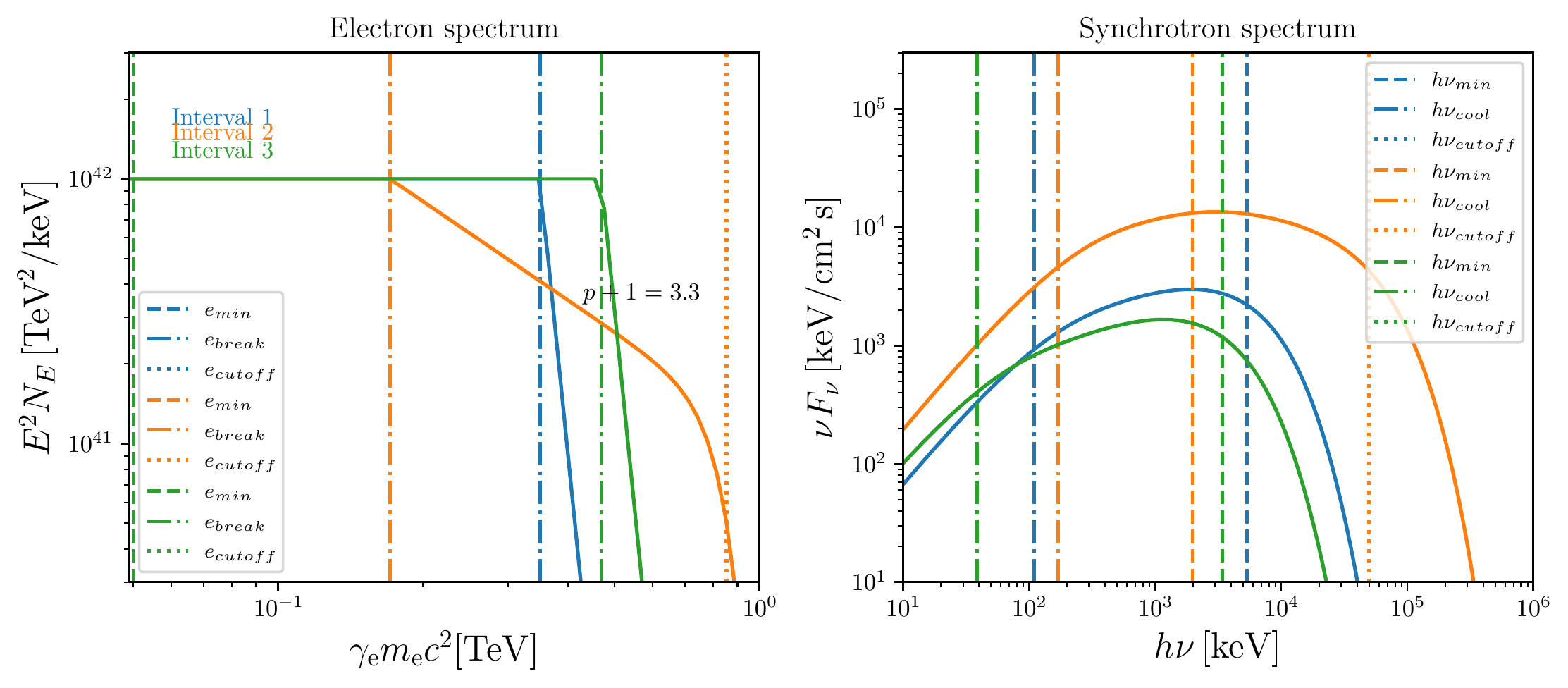}
\caption{{Spectral distributions of electrons and gamma-rays.} Left panel: Electron energy spectrum used in fits to the three time intervals. Since the fits are translationally degenerate, the energy scale is arbitrary. The physical quantities are later derived in Appendix \ref{app:12}. Right panel: $\nu F_{\nu}$-spectrum from the three time intervals, being the best fits to the data.  
%Fit to the three intervals of the main episode of GRB~160821A. Note that the electron distribution is normalised... 
}\label{fig:1}
\end{centering}
\end{figure*}

\vspace{5mm}
 \begin{table*}
	\centering
		\begin{tabular}{*{6}{l}}
		\hline
		 Time intervals &
		 $h \nu_c$ & $h \nu_m$ & $h \nu_{cutoff}$&
		 $F_{\nu} (h \nu_c)$   & $p+1$\\
		 & (keV) & (keV) & (keV) & (cm$^{-2}$ s$^{-1}$) & \\
		 %&\multicolumn{1}{c|}{$p$} & \multicolumn{1}{c|}{$\Gamma_0$}\\
		 %& & & & & & & \multicolumn{2}{c|}{(p is free)} &   & & \\
		%\cline{2-7}
		 %& AIC & AIC  & AIC & AIC  & AIC & \\
		\hline
		\\
		Interval 1 &  $108.3^{+11.8}_{-12.3}$ & $5.5^{+1.0}_{-0.9}\,\times\,10^{3}$ & &$8.5^{+0.1}_{-0.1}$ & \\
		\\
		Interval 2 &$167.4^{+14.2}_{-12.4}$ &$2.3^{+1.1}_{-1.0}\,\times\,10^{3}$& $5.2^{+1.9}_{-1.7}\,\times\,10^{4}$ &$27.2^{+0.3}_{-0.4}$ &$3.3\pm 0.2$ \\
		\\
		Interval 3 &$38.8^{+5.2}_{-5.6}$ &$3.4^{+1.1}_{-0.9}\,\times\,10^{3}$&  &$10.4^{+0.2}_{-0.2}$ & \\
		\hline
		
	\end{tabular}
	\caption{Measured spectral properties using the synchrotron emission model. The values correspond to the means of the respective marginalised posterior distributions. The intervals denote Bayesian credible intervals corresponding to the 95\% highest density interval.}
\label{tab:1} 
    \end{table*}

\section{Scenario derived from the observations}
\label{sec:scenario}
Since the flare is very bright it has to have originated in an external shock, as any internal shocks are too inefficient \citep{Kobayashi_etal_1997, Beloborodov2000, Spada2000, Kobayashi2001}. Furthermore, it cannot either be a collision between a late emitted shell that catches up a shell from an early ejection by the GRB. At the collision time of $\sim 100$s both such shells have to be relativistic and the relative Lorentz factor between  the colliding shells would therefore be low. A large 
contrast in Lorentz factors is needed to explain the exceptional brightness of the flare. 
%%%%%%
%%%%%%
Moreover, the main emission episode cannot be part of the long-lived, self-similar afterglow emission.
The reason is that an additional high-energy component emerges at the end of  the main emission episode and appears as a separate spectral component at around 185\,s \citep{Sharma_etal_2019}. This component is strongly required by the data and  gives a change in AIC of 116. In addition, significant emission above $100$ MeV from this component was observed by Large Area Telescope (LAT) onboard {\it Fermi} for a period of $\sim$ 2000s after the Gamma-ray burst monitor (GBM) trigger time ($T_0$),  decaying as a power-law in time with the temporal index of $1.15 \pm 0.10$ in the Fermi/LAT energy range.   The additional, long-lived component is thus naturally related to the afterglow, produced by an external shock in the self-similar blastwave regime \citep{Ajello2019}.
The onset of the afterglow emission at $t \sim 185$s, indicates a deceleration radius that is larger than the emission radius  of the main episode, which peaks at 135 s. %of $r_{\rm dec} > 10^{17}$s. 
%As mentioned above, the high efficiency requires that the late central engine activity 
The conclusion is therefore that the main emission episode must be due to a blastwave that 
encounters a dense circumburst shell that is, at most, mildly relativistic, lying within the deceleration radius. %, since only the differential kinetic energy can be extracted, which needs to be large. 
 As this encounter occurs already at $\sim$ 130s, such a shell therefore needs to be a preexisting structure, since any earlier GRB ejection would still be relativistic at that time. A plausible origin of such a pre-existing structure are the ring nebulae around the progenitor Wolf-Rayet stars (WR). These nebula are {either} caused by massive winds, which sweep up the circumstellar medium, {or by
 instabilities that cause elevations of the outer envelope leading to occasional giant eruption events, with major mass ejections \citep{Chu81, Crowther2007}.  
 Such events are thought to} cause narrow, nearly spherical shells around the progenitor star %, sometimes with complex structures
\citep{JH1965}.  Up to a third of  WR stars observed in the Galaxy have a narrow ring nebula \citep{Marston1997} lying at a typical distance of 1 pc from the central star, and some having much smaller sizes \citep{Stock2010}.  The existence of a low-density cavity within such wind-blown bubbles \citep{Toala2013} would lead to very little interaction with the blast wave before it encounters the circumstellar ring itself. This fact is supported by the quiescent period observed just before the 130 s flare in GRB160821A.

Finally, the ratio of pulse-width to pulse-time for the main episode $40\, \mathrm{s} / 135\, \mathrm{s} < 1$, which indicates that it is due to late central engine activity \citep{Lazzati_Perna2007, Pereyra2021} causing an external-shock flare.  
This is reminiscent of strong gamma-ray flares observed at the end of the prompt phase \citep{Zhang_etal_2018_160625B} %(Zhang, B.B. et al. 2017) 
and X-ray flares observed after the prompt activity in the gamma-ray band \citep{Hu_etal_2014}, 
which all require a long-lived central engine activity.

Within this scenario the synchrotron spectra from the analysis above can be translated into physical properties of the plasma. {Since both the bulk Lorentz factor and the redshift  are unknown,  we initially use the fiducial value of $\Gamma =300$ and  the estimated value of $z=0.4$ (see Appendix \ref{app:115})}. The physical properties  are  derived in Appendix \ref{app:12}. { Table~\ref{tab:2} gives the derived values of $\gamma_{\rm c}$,  $\gamma_{\rm m}$, $B$, the emission (dynamical) radius $R_{\rm dyn} = 2c \, \Gamma^2\, \Delta t$, where $\Delta t$ is the pulse duration, total number of radiating electrons $N_{\rm e}$, and finally the magnetisation, $\sigma = B'^2/ 4 \pi \Gamma  n' \mpp c^2$. Here, the primed quantities are in the comoving frame and $n'$ is the particle density. The magnetisation is, therefore, determined for the downstream of the shock}. 
The typical Lorentz factor of the electrons is found to be very high $\gamma \sim 10^{5}$. Its value is given by $\gamma \sim \epsilon_{\rm e} (\mpp /\me) \Gamma$, where $\Gamma$ is relative Lorentz factor between the colliding shells, $\mpp$ and $\me$ are the proton and electron masses and $\epsilon_{\rm e}$ is the energy partition fraction.  The high value of $\gamma$ is therefore consistent with the external shock scenario and a large contrast in Lorentz factors.

There are three main changes of the physical properties between the first and second episode:
(i) The number of emitting particles, $N_{\rm e}$, increases by a factor of 4.
(ii) There is an onset of particle acceleration to a power-law with $p=2.3$, which contains around 10\% of the particles. (iii) The magnetisation $\sigma$ decreases.  At the same time, the $B$-field does not change very much, while $\gm$ and $\gc$ are the same to within a factor of 2. We note that there is a small but significant decrease  in $\gm$ by 30\%.
{From a theoretical point-of-view, both the fraction of particles that have been accelerated in Interval 2 ($\sim  10 \%$) and the power law slope of the injected electrons ($p\sim 2.3$) are in line with the robust expectations for particle acceleration in weakly magnetised flows \citep{Sironi2015}.  The high-energy cutoff during Interval 2 is at around 50 MeV (Tab. \ref{tab:1}). Such a cutoff is expected from shock acceleration and depends on many factors, such as the shock duration \citep{Kirk2010, Sironi2013} and magnetic field configuration \citep{Lemoine2013}.

The observed increase in emitting particles during Interval 2 can have different reasons. 
In the encounter between the blastwave and the slow moving and dense pre-existing shell, relativistic forward and reverese shocks will develop, leading to two distinct emission zones.  The properties of these shocks are expected to be similar, since the slow shell is pre-existing \citep{PeerLC2017}.   The forward shock is pronounced during Interval 2, as it moves through the denser shell, and accelerates its particles across the shock into a power law distribution. Intervals 1 and 3 are then related to reverse shock moving into the blastwave and heating its particle content. 
Alternatively, the observed emission is only from the forward shock that encounters fluctuation in the particle 
density in the preexisting shell.  
In both cases, the denser regions causes larger release of energy leading to the change  of the observed intensity. Moreover, the increase in particle density
%number by a factor of five is expected to decrease $\gm$ by a factor of around two\cite{Peerwijers2006}, which is largely similar to what is observed.
is  expected to slightly decrease $\gm$ \citep{Peerwijers2006}. This is largely similar to what is observed.
Finally, as the $B$-fields are relatively constant, the increase in emitting particles also explains the observed  decrease in magnetisation, $\sigma$.
}

\vskip 10mm

\section{Onset of particle acceleration}
The onset of particle acceleration can therefore naturally be related to the variation in magnetisation. 
We find that $\sigma$ in GRB~160821A decreases by a factor of 5 in interval 2 and then increases again by a factor of 10.
It might thus have dropped below a critical value for acceleration to become operative during interval 2.  One possibility, 
that is directly related to the magnetisation at a relativistic shock, is the finding that the microturbulence centers (e.g. caused by Weibel instabilities) needed for Fermi acceleration to operate, cannot be formed if the magnetisation exceeds a certain  critical value \citep{, LemoinePelletier2011, Lemoine2013}. Analytical work and numerical simulations \citep{Sironi2011, Lemoine_etal_2013, Pelletier_etal_2017} show that the {theoretical value of the critical magnetisation is $\sigma_{\rm c} \sim 10^{-6}$ (assuming $\Gamma = 300$). %\, \Gamma_{\rm r}^2$, where $\Gamma_{\rm r}$ is the relative Lorentz factor across the shock.  The value of $\sigma_{\rm c}$ is, however, 
This value is, however, consistently larger than what is found during the main episode in GRB~160821A, which has an magnetisation $\sigma \sim 10^{-7}$ (Tab. \ref{tab:2}), indicating that particle acceleration ought to appear throughout its duration.} %, where we have assumed that the magnetic field is oriented predominantly transverse to the shock normal \citep{Medvedev&Loeb1999, Frederiksen2004}.
On the other hand, as shown in Appendix \ref{app:2} a viable solution\footnote{A comparison between Tables \ref{tab:2} and \ref{tab:4} also illustrates the sensitivity of the derived parameter values to the assumed value of the Lorentz factor, $\Gamma$. Most of the derived parameters only weakly depend on $z$ (see app. \ref{app:12}), apart from $N_e$ and $\sigma$, which vary around a factor of five between $z=0.4$ and $z=1.0$.
} exists for larger values of the Lorentz factor, $420  \simleq \Gamma  \simleq 770$,
%which though are values that are larger than expected in the GRB standard sce 
which corresponds to shock radii at $R\sim 10^{18}$ cm. {In such a case, the magnetisation drops below the critical value as stipulated.}
{We note, however, that such large values of $\Gamma$ and $R$ require very low particle densities of the  circumburst medium \citep[e.g., ][]{ReesMeszaros1992}. With the averaged value of $\Gamma = 595$ from the range above, the particle density required is as low as $ \sim  10^{-3}$ cm$^{-3}$, which indicates a tenuous gas of the immediate burst surrounding.}
On the other hand, $R = 10^{18}$ cm is the typical size of WR ring nebulae {which, combined with the low densities of their interiors,} makes the large values of $R$ and $\Gamma$ consistent with the scenario presented above in section \ref{sec:scenario}. %We note that similarly large values of $\Gamma$ and $R$ are typically found for synchrotron spectral fits to GRB prompt data \citep{Kumar&McMahon2008, burgess2020}. 
If indeed the onset of particle acceleration is  caused by Fermi acceleration and determined by the survival of microturbulent magnetic fields \citep{Lemoine_Pelletier2010}, then the observed transition from acceleration to non-acceleration gives a new, independent way to constrain the bulk Lorentz factor. % and that $\Gamma_{\rm r}$ is large. 

{Another possibility for the observed particle acceleration is the converter acceleration mechanism \citep{Derishev2003}. In this mechanism, $e^{\pm}$ pairs in the upstream gain energy as they cross the shock front. The energetic electrons then cool by inverse Compton emission in the downstream, producing energetic photons, which can propagate back across the shock to the upstream. If the opacity for photon-photon interaction is high enough, $e^{\pm}$ pairs can again be created in the upstream, thereby completing a  Fermi cycle.   The converter mechanism is very efficient \citep{Derishev2016} but there are many mechanisms that can counteract it \citep{Derishev2017}. For instance, the relative efficiency of inverse Compton emission  will change depending on the magnetic field strength, which can lead to synchrotron losses to become dominant. A change in the Lorentz factor jump across the shock front will also affect its efficiency.
If the powerlaw distribution of electrons observed in GRB160821A is due to the converter mechanism, then the onset of particle acceleration must be caused by an increase in its efficiency, causing it to become operational. Neither a change in magnetic field strength nor a change in Lorentz factor jump is, however,  expected in the scenario described above in section \ref{sec:scenario}.}
%The observed onset of particle acceleration indicates, in such a scenario, that the converter 

\begin{table*}
	\centering
		\begin{tabular}{*{8}{l}}
%\begin{tabular}{|c|c|c|c|c|c|c|c|c|c|}
	% four columns, alignment for each
		\hline
		 Time intervals &
		 $\gamma_{\rm c}$ & $\gamma_{\rm m}$ & $B$ &
		 $R_{\rm dyn}$ &  $N_{e}$  & $\sigma$\\ 
		 & ($10^{5}$) & ($10^{5}$) & (Gauss) & ($10^{17}$ cm) & ($10^{49}$) &  ($10^{-7}$)\\ 
		 \hline
		 \\
		 Interval 1 & $1.33^{+0.09}_{-0.10} $ & $9.4^{+0.8}_{-0.8}$ & $1.65^{+0.06}_{-0.06}$ & $2.2$ & $4.33^{+0.09}_{-0.09}$  & $3.9^{+0.3}_{-0.4} $ \\ 
		 \\		 
		 Interval 2 & $1.78^{+0.10}_{-0.09}$ & $6.5^{+1.2}_{-1.2}$ & $1.43^{+0.03}_{-0.04}$ & " & $16.1^{+0.2}_{-0.2}$  & $0.78^{+0.05}_{-0.05}$ \\ 
        \\
		 Interval 3 & $0.67^{+0.06}_{-0.07}$ & $6.3^{+0.8}_{-0.8}$ & $2.32^{+0.12}_{-0.09}$ & "  & $3.78^{+0.10}_{-0.11}$   & $8.8^{+1.1}_{-1.0}$  \\
	\hline
	\end{tabular}
	\caption{Derived physical parameters, based on the assumption that the observer frame cooling time, $t_{\rm cool}$ is the same for all intervals. The fiducial value of $\Gamma= 300$ and $z=0.4$ are assumed. %The values presented here are derived further assuming a bulk Lorentz factor, $\Gamma = 100$ and redshift, $z=1$. 
	The intervals denote Bayesian credible intervals corresponding to the 95\% highest density interval. %Note $R$  in interval 2 and 3 are set to be the same as in interval 1.
	}
\label{tab:2} 
\end{table*}

\section{Discussion and conclusion}

We have analysed the gamma-rays in GRB160821A, which can be convincingly identified as synchrotron emission.
%We have shown that correct identification 
Synchrotron modelling of the observed data consequently reveals the energy distribution of the radiating electrons. We find for the first time evidence for the onset of the acceleration process, in which a fraction of thermally distributed particles are accelerated into a power law distribution to higher energies. We argued that the strong flare in GRB160821A is due to an interaction between a late blastwave interacting with preexisting shell of slowly moving material, such as a Wolf-Rayet ring nebula. This causes a forward and reverse shock, that both are relativistic and have similar properties \citep{PeerLC2017}.
The particle acceleration detected in interval 2, occurs in the forward shock, which encounters a denser region
 %process starts when the emission intensity is the largest and that it is accompanied by a swing in the polarisation angle and degree. %We interprt 
and a lower magnetisation. %: Even though the $B$-field is observed to be similar through-out the emission episode, the larger density of electrons in the preexisting shell causes a smaller magnetisation, $\sigma$, in interval 2.  
%This new observational constraint on the comoving magnetisation of the plasma can be used to determine additional quantities of the GRB, such as the bulk Lorentz factor, and be compared to theories of particle acceleration in collisionless astrophysical shocks. 

The high degree of polarisation observed during the main episode indicates that the jet should carry a dominant ordered magnetic field component in a scale larger than $\Gamma^{-1}$ or a globally ordered toroidal field \citep{Sharma_etal_2019}.
In addition, a consequence of particle acceleration is that a shock-generated, small-scale, random $B$-field is formed \citep{Keshet2009}. Therefore, such a field should exist in addition to the ordered field during Interval 2 in GRB~160821A. The combination of these field components can change the resulting weight of the polarisation contributions over the jet image, thereby altering the observed polarisation degree and angle \citep{Granot2003, Gill2021, Lan2020}. 
This could be the explanation to the fact that the polarisation angle was found to change twice, first by around $80^{\circ}$ 
and later back again to its original value in GRB160821A \citep{Sharma_etal_2019}.
We note that the polarisation degree indeed decreases during Interval 2, even though the errors are large on the measurements.

A consequence of the results presented in this paper is that only a faction of GRBs should have bright, late synchrotron pulses, since nebula rings are only observed to occur in a fraction of Wolf-Rayet stars. Such late prompt synchrotron emission provides diagnostics of  the inner parts of the progenitor winds, that were emitted a few centuries prior to the GRB explosion.

\section*{Acknowledgements} 
{We thank the anonymous referee for useful suggestions and Dr. Filip Samuelsson for useful discussions}. We acknowledge support from the Swedish National Space Agency (131/18 and 2020-00084) and the Swedish Research Council (Vetenskapsr{\aa}det, 2018-03513, 2020-00540). S.I. is supported by DST INSPIRE Faculty Scheme (IFA19-PH245). This research made use of the High Energy Astrophysics Science Archive Research Center (HEASARC) Online Service at the NASA/Goddard Space Flight Center (GSFC). In particular, we thank the GBM team for providing the tools and data.

%\bibliography{ref2020.bib}
\bibliography{ref2020.bib}

%\newpage
\appendix

\section{Synchrotron Modelling} 
\label{app:1}
We model the emission as fast-cooled synchrotron emission, as described in \citet{Aharonian2010}. Specifically, we use the {\it Naima} software package \citep{Naima2015} to carry out the calculations of model spectra. The distribution of  electrons with energy $e$ is modelled by a broken powerlaw, with a low-energy ($e < e_{break}$) slope fixed at $\alpha = 2$ (expected for fast cooling synchrotron). In the presence of a population of accelerated electrons, a high energy power law ($e > e_{break}$)
 is free to vary, and includes an exponential cutoff at high energies, $e_{cutoff}$.
For intervals 1 and 3 our initial fits yield posteriors of $p > 10$ and $e_{cutoff}$ tending to merge with $e_{break}$. These results imply that there is no high-energy powerlaw (p) and a $e_{cutoff} > e_{break}$ is not found. Physically, this means that there is no evidence that particle acceleration has taken place and that the electrons from a heated quasi-Maxwellian distribution. In order to represent such a very narrow electron distribution, we choose to freeze $p$ and $e_{cutoff}$ at large values in intervals 1 and 3.  This in turn allows us to get better constraints on all other parameters.
We note that after the convolution with the synchrotron kernel the observed spectra from a quasi-Maxwellian distribution and our simplified electron distribution  will be indistinguishable. In Table~\ref{Table:fix_params} we present the values of all frozen parameters of our model. All fixed parameters are frozen to the same values across the three intervals, with the exception of $p$ and $e_{\rm cutoff}$, which are free in interval 2.   Note that we fit for breaks in the electron spectrum, mainly $e_{\rm break}$, which is more commonly parameterised in terms of the comoving electron Lorentz factor, $\gamma$. This, in turn, can be translated to where the break lies in the photon spectrum, $h\nu$.

\begin{table}[]
    \centering
    \begin{tabular}{c|c|l|l}
    \hline
         Parameter & Fixed Values & Unit & Description\\
         \hline
         $N_{\rm e,0}$ & $10^{42}$ & keV$^{-1}$ & Amplitude at the $e_{break}$ of the electron distribution \\
         $e_0$ & $1$ & TeV & Reference point for electron distribution broken powerlaw  \\
         $\beta$ & 10 &  & Sharpness of high energy cutoff in the electron distribution\\
         $\rm e_{min}$ & 0.05 & TeV & Minimum electron energy for the electron distribution\\
         $\rm e_{max}$ & 1000 & TeV & Maximum electron energy for the electron distribution \\
         nEed & 10 &  & Number of points per decade in energy for the electron energy \\
         & & & and distribution arrays \\
         $\alpha$ & 2 &  & Power law index for $e < e_{break}$\\
         %$\gamma_{\rm cool}$ & $10^{4}$ & \\
         $p$ $^{*}$ & 20 & & Power law index for $ e > e_{break}$ \\
         $e_{\rm cutoff}$ $^{*}$ & $50$ & TeV & Cutoff energy at higher energies of the electron distribution \\
         $e_{break}$ &  & TeV & Break energy of electron distribution broken powerlaw\\
         $B$ & & Gauss & Isotropic magnetic field strength\\
         norm & & & Normalization (differential flux at a distance of 1 Mpc) \\
         
         \hline
    \end{tabular}
    \caption{List of parameters of the synchrotron model. The values of the parameters which are kept fixed are given. 
    %List of fixed parameters and the values at which they are frozen. 
    Note that $^{*}$ $p$ and $e_{\rm cutoff}$ are free parameters in interval 2. 
%%%%%%%%%%%%%%%%%%%%
%%%%%%%%%%%%%%%%%%%%%%%
    }
    \label{Table:fix_params}
\end{table}

\subsection{Spectral analysis}
The spectral analysis is carried out in the Multi-Mission Maximum Likelihood (3ML) software \citep{Vianello2015}. We implement a Bayesian analysis, in which we evaluate the posterior of our model conditioned on observed data using MultiNest \citep{Multinest2011} implemented in python \citep{Buchner2016}. The analysis is carried out using 1000 live points.

The priors for the free parameters in Intervals 1 and 3 are given by
\begin{align*}
    P(e_{\rm break}) &= U(0.05,30)~({\rm TeV}) \\
    \log P(B) &= U(10^{-1},10^{4})~({\rm G}) \\
    \log P({\rm norm}) &= U(10^{-2},10^{2}).
\end{align*}
In addition to these parameters, in Interval 2 we also fit for $p$ and $e_{\rm cutoff}$, which are given the following priors
\begin{align*}
    P(p) &= U(2,20) \\
    \log P(e_{\rm cutoff}) &= U(10^{-1},10^{1})~({\rm TeV}).
\end{align*}

Further, the analysis was carried out using different priors and with different numbers of live points (500 and 2000). From trying both wider and more narrow priors on all parameters, we find no significant impact on the results, as long as the priors include the mode of the posterior with some margin. Using 500 live points was sometimes sufficient for convergence, whereas we found no significant difference between using 1000 or 2000 live points. We thus conclude that our results are not sensitive to our choices of priors, and further our posteriors are sampled satisfactorily.

\begin{figure}[ht!]
\centering
\includegraphics[width=0.48\columnwidth]{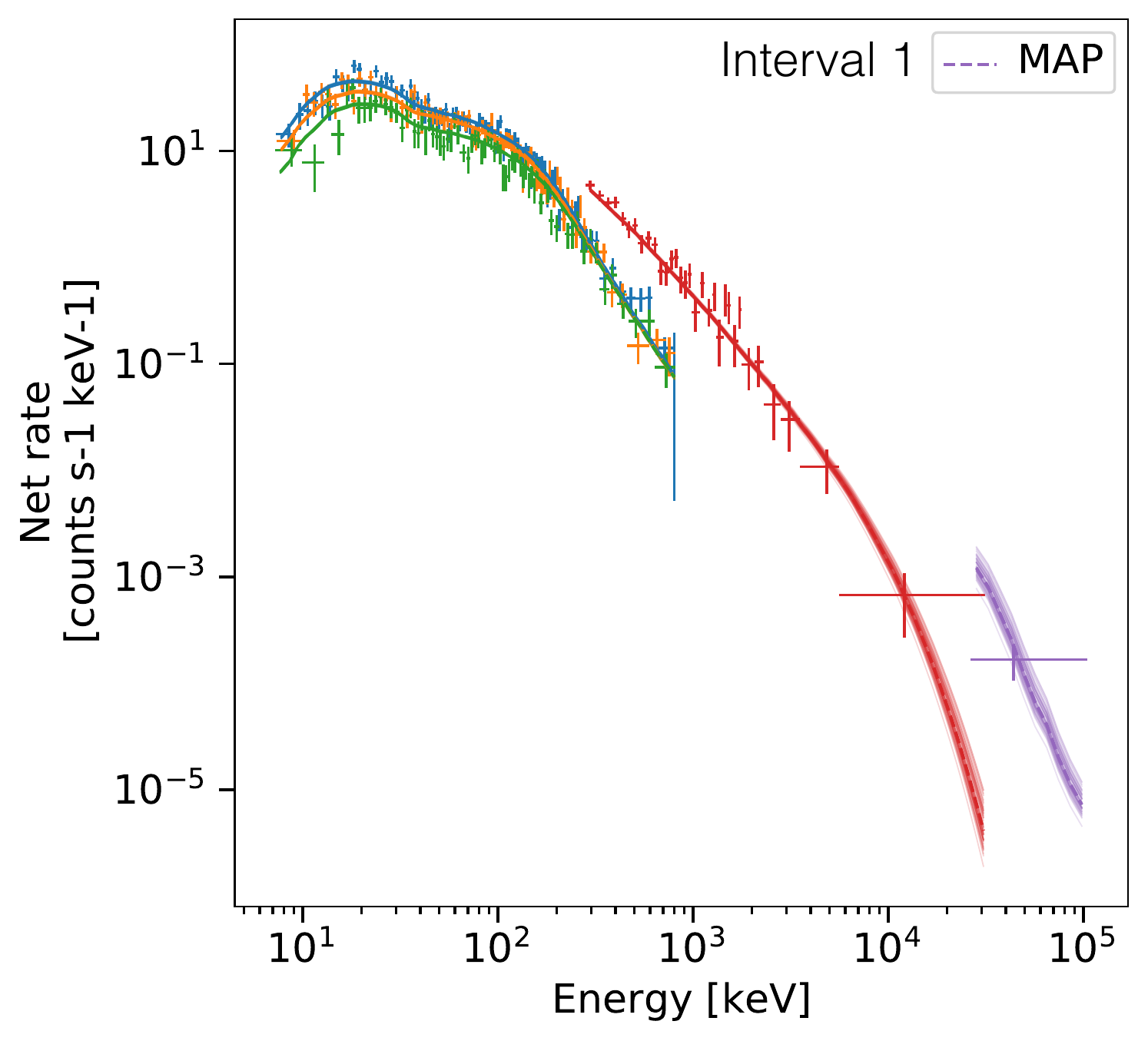}
\includegraphics[width=0.48\columnwidth]{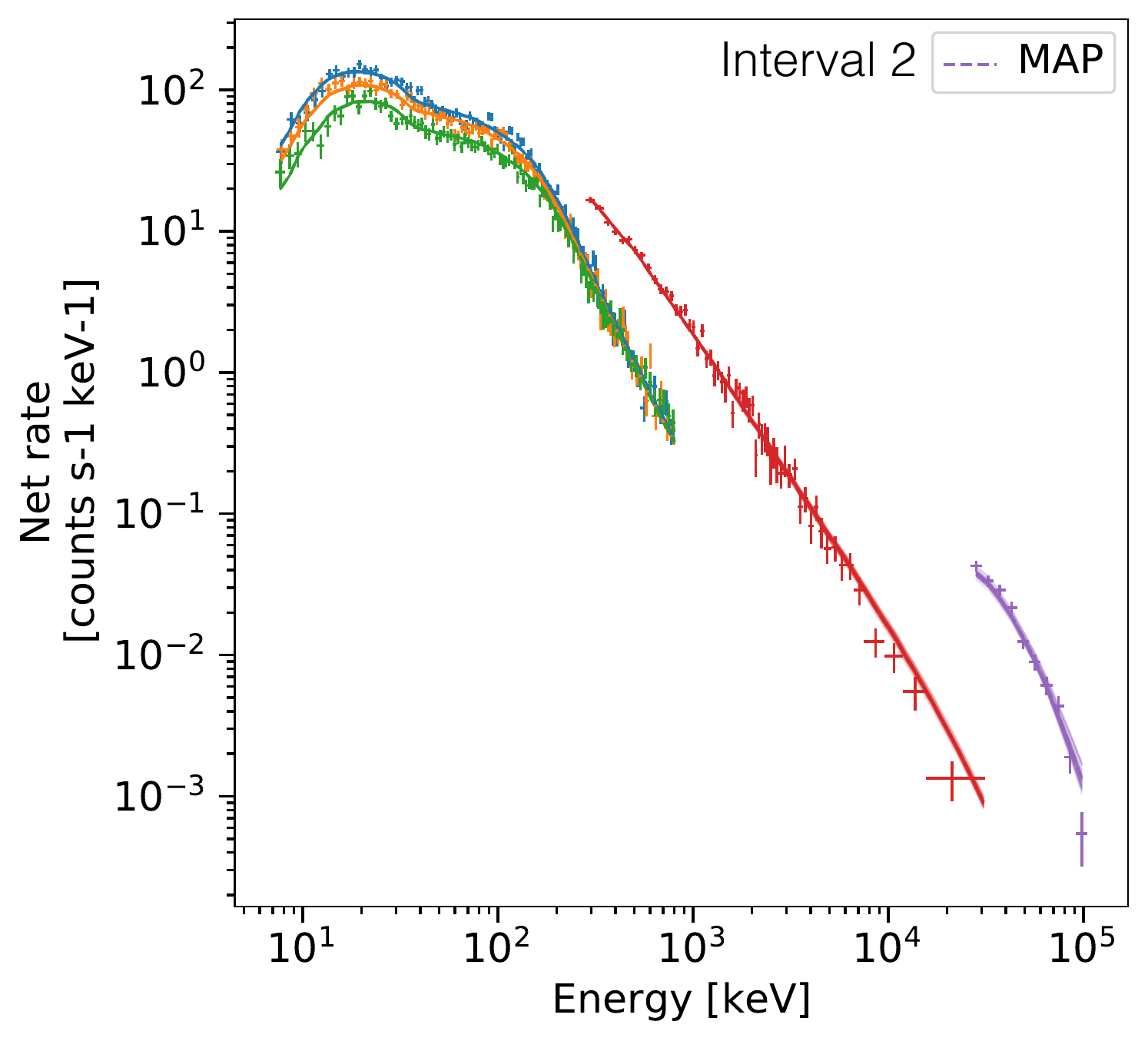}
\includegraphics[width=0.48\columnwidth]{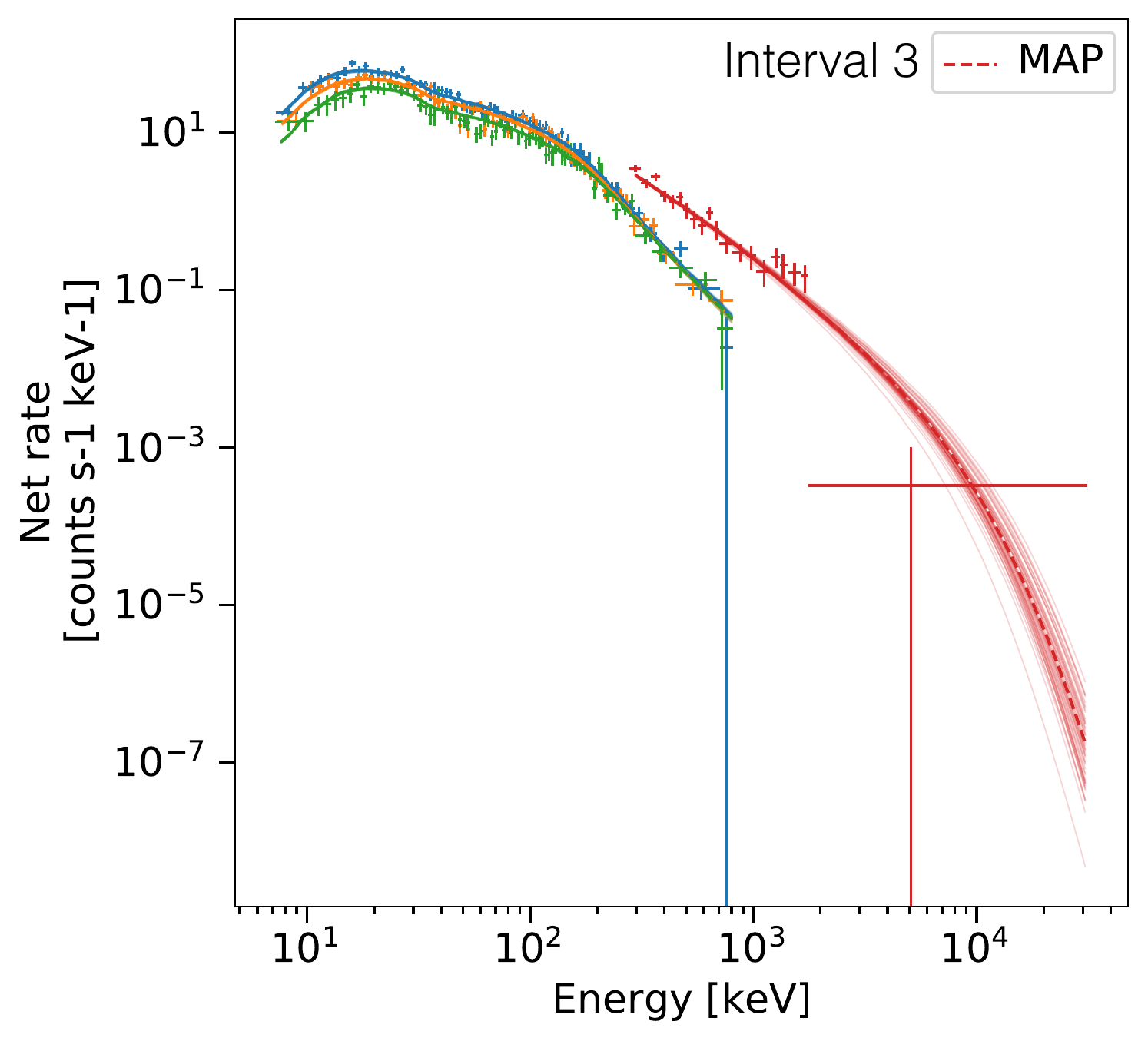}
\caption{Draws from the posterior distribution of the model conditioned on the data in the three intervals. The 50 thin lines correspond to random draws from the posterior distribution and the dashed line corresponds to the MAP. The data and model have been re-binned to bins containing at least 20 counts. The blue, orange, green, red, and purple colours correspond to the NaI6, NaI7, NaI9, BGO1, and LAT/LLE data, respectively.}
\label{fig:fits_wMAP}
\end{figure}

In Fig.~\ref{fig:fits_wMAP} we present fits of the model to the data in terms of draws from the posterior distribution plotted together with the observed data in count space. Additionally, the plots also contain the maximum a posteriori (MAP) estimate, corresponding to the mode of the posterior distribution. By visual inspection there is a decent agreement between the model and the observed data. 
{In the third interval, there was no significant LAT-LLE data.  In the LLE energy range, the model predicts a flux well below the detection threshold, consistent with the observations. In order to reach 
a detection significance of 4$\sigma$ \citep[the threshold used by the Fermi-LAT collaboration, e.g., ][]{LAT2019}), the model flux in this energy range would need to be doubled.}

% \begin{figure}[ht!]
% \centering
% \includegraphics[width=0.8\columnwidth]{figures/posterior_corner_plot_original_parameters_v38_allIntervals.pdf}
% \caption{Corner plot of posterior probability distribution of the free parameters of our model. The blue, green, and red colours correspond to interval 1, 2, and 3, respectively.}
% \label{fig:cornerplot_small}
% \end{figure}

\begin{figure}[ht!]
\centering
\includegraphics[width=0.8\columnwidth]{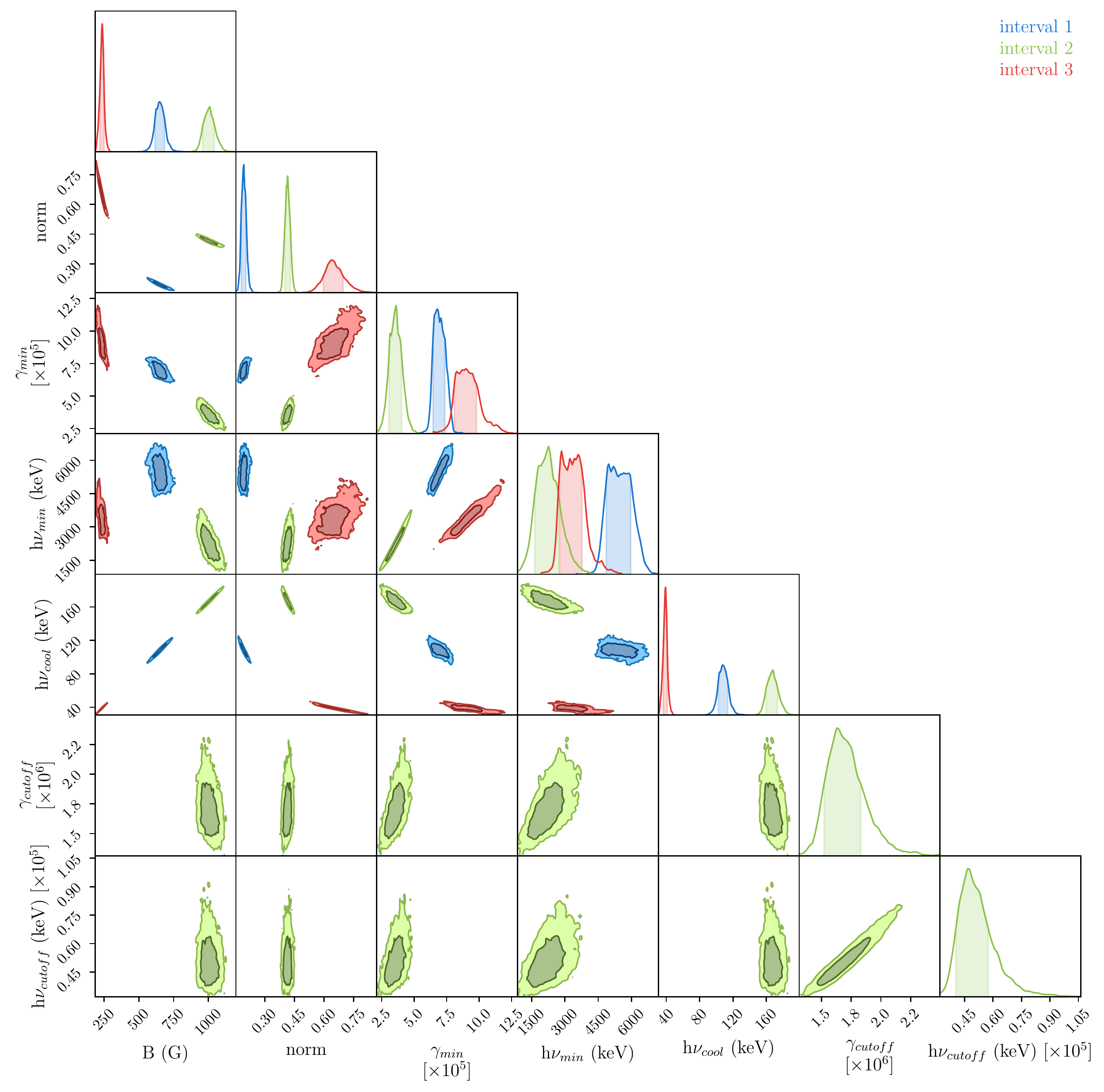}
\caption{Corner plot of posterior probability distribution of free model parameters, as well as the derived values of the break energies in the photon spectra. The blue, green, and red colours correspond to Interval 1, 2, and 3, respectively, and the light and dark contours represent the 68\% and 95\% credible regions.}
\label{fig:cornerplot_large}
\end{figure}

\begin{figure}[ht!]
\centering
\includegraphics[width=0.5\columnwidth]{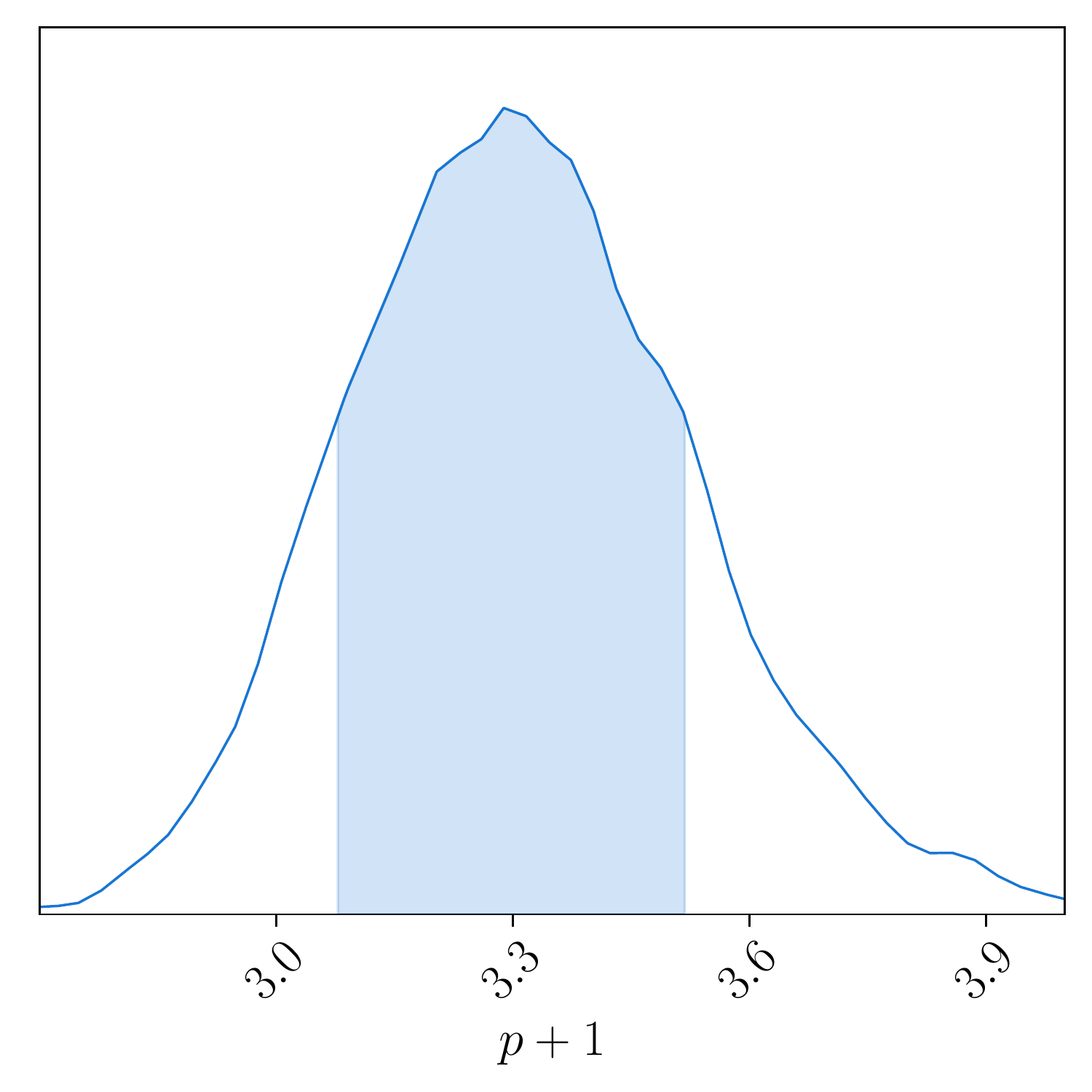}
\caption{The posterior distribution of the fitted power-law index of the accelerated electron distribution in Interval 2. The injected power-law slope $p$ is steepened by unity due to the cooling. The shaded area is the 95\% credible region. The fitted value of $p=2.3 \pm 0.2$ is in line with the robust expectations for particle acceleration in weakly magnetised flows \cite{Sironi2015}.}
\label{fig:Interval2_p}
\end{figure}

Figure~\ref{fig:cornerplot_large} shows the posteriors for the three intervals together. We include all free parameters in our analyses. Finally, Figure~\ref{fig:Interval2_p} shows the posterior distribution of the fitted power-law index of the accelerated electron distribution in Interval 2. The injected power-law slope $p$ is steepened by unity due to the cooling. The fitted value is therefore $p=2.3 \pm 0.2$.

 The emergence of a high-energy flux distribution in Interval 2 is very significant.
 We find highly significant changes in information criteria AIC and BIC between a fit with and a fit without a high-energy powerlaw distribution of the electrons:   $\Delta$AIC = 102 and $\Delta$ BIC = 95. In this comparison the parameters $B$, $e_{break}$ and $norm$ were free to vary. In addition, we also find that the residuals between the best fit model and the data have a very pronounced wavy structure, which is indicative for the need for an additional component in the spectral model.

%\vskip5mm
%\noindent{\bf Derivation of physical parameters from the synchrotron fits.} 
\begin{comment}
The observed cooling break in interval 1 is given by
\begin{equation}
\nuc_1 = \frac{3 q_{\rm e} h}{4\pi m_{\rm e} c (1+z)} \, \gb \gc_1^2 B_1
\label{eq:1}
\end{equation}
and the associated cooling time 
\begin{equation}
t_{\rm cool} = \frac{6 \pi \me c}{\sigma_{\rm T} \gc B^2 \gb} = 10 \mathrm{s} 
\label{eq:cool}
\end{equation}
is equated to the timebin size of 10s. Since the spectra are translationally degenerate, we haev to make assumptions on the unknown redshift, which is set to $z=1$, (see Methods, 'Broad band emission and redshift') and, e.g., the value of $\Gamma_1$, which is set at $\Gamma_1 = 2$. We can then solve for 
the value of $B_1$ and $\gb$ and find the values of  $\gc= \epsilon_{\rel e} m_{\rm p}/ m_{\rm e} \grel   = 1.3 \times 10^5$, and \Gamma_{2} = \gb.
%can 

Next, the measured ratio of 
\begin{equation}
{\frac{\nuc}{\num}} = \left( \frac{\gc}{\gm} \right)^2
\end{equation}
which gives a value of $\gm_1$.
\end{comment}

\subsection{Estimation of the redshift}
\label{app:115}

{ The redshift of GRB160821A is not known. While the determined spectral shape and the identification of the onset of particle acceleration is independent of this fact, the determined energetics and magnetisations will depend on the assumed value. We will assume the value to be $z = 0.4$ based on the following reasoning.

The fluence of GRB160821A in the range 10 keV - 5 GeV is $(1.30 \pm 0.03) \times 10^{-3}$ erg cm$^{-2}$, which makes it among the brightest observed \citep{Sharma2019}. This means that the burst is either very close or, if distant, exceptionally bright.  GRBs have a broad distribution of isotropically equivalent energy output, $E_{\rm iso}$, reaching up to $\sim 7 \times 10^{54}$ erg for GRB160625B \citep{Sharma_etal_2021}. Assuming $E_{\rm iso} = 7 \times 10^{54}$ erg for GRB160821A yields an upper value of the redshift to a modest $z = 1$.  On the other hand, there was no associated supernova detection for GRB160821A. In addition, we searched various optical surveys such as Sloan Digital Sky Survey (SDSS), Panoramic Survey Telescope and Rapid Response System (Pan-STARRS) as well as radio source surveys of both VLA and GMRT, from the database maintained by VizieR Catalogue Service, and did not find any evidence for host galaxy at the GRB localization. This led us to the inference that the host galaxy in optical is very faint (R-mag > 23.2, \citealt{Panstarrs1_etal_2016}). Following the study of the host galaxies of long GRBs \citep{Hjorth_etal_2012,Jakobsson_etal_2012}, the lowest redshift at which faint long GRB host galaxy with an R-mag > 23 that has been detected till date is around $z=0.4$. Therefore,  $z=0.4$ serves as a lower limit for the redshift in GRB160821A.

Two further arguments support the lower value of the redshift. First, the highest energy of the LAT photons calculated in the rest-fame for GRB160821A is $E^{\rm LAT}_{\rm max}\sim 4.7 (1 + z)$ GeV $< 10$ GeV, which is among the lowest $20\%$ of observed  $E^{\rm LAT}_{\rm max}$-values \citep{Ajello2019}. 
Since there is an observed positive trend between $E^{\rm LAT}_{\rm max}$ and $E_{\rm iso}$ \citep{Ajello2019}, 
the low value of $E^{\rm LAT}_{\rm max}$ in GRB160821A indicates a low value of $E_{\rm iso}$. 
Second, the late onset time of  the afterglow emission ($\sim 185$ s) indicates again that the $E_{\rm iso}$ should be relatively low  \citep{Ghirlanda2018}. Low $E_{\rm iso}$ combined with the measured fluence supports that GRB160821A is not very distant.     

We, therefore, use the estimate of the redshift $z=0.4$, which, for GRB160821A, implies a value of $E_{\rm iso} = 7.6 \times 10^{53}$ %6.9 \times 10^{53}$ 
erg in the gamma-rays\footnote{The isotropic burst energy estimate in this Letter uses the standard $\lambda$CDM cosmology, with cosmological parameters, $H_0 = 67.4 \pm 0.5$ km s$^{-1}$ Mpc$^{-1}$, $\Omega_{vac} = 0.685$ and $\Omega_{m} = 0.315$ \citep{Planck2020}.}. Such a value is within the typical range of other LAT detected GRBs \citep{Racusin2011, Ajello2019}. }

\subsection{Derivation of physical parameters}
\label{app:12}
We will assume the bulk Lorentz factor, $\Gamma= 300$ and {the redshift to be $z=0.4$}. In order to estimate values of the magnetic field and the characteristic Lorentz factors of the electrons, we first use the value of the observed cooling break in interval 1 which is given by
\begin{equation}
{\nuc}_1' = \frac{3 q_{\rm e} h}{4\pi m_{\rm e} c } \,  {\gc}_1^2 B_1 = h\nu_{\rm c1}^{\rm obs}\, \Gamma^{-1} (1+z)
\label{eq:1}
\end{equation}
%Therefore, with the above assumptions on $\Gamma$, $\gc_1$, and $z$ we can get a value of $B_1$. 

We then set the cooling time to be the width of the timebin, $t_{\rm bin} \sim 10$s:
\begin{equation}
t_{\rm c}' = \frac{6 \pi \me c}{\sigma_{\rm T}} \gc^{-1} B^{-2} =  t^{\rm obs}_{\rm c} \, \Gamma (1+z)^{-1}\sim 10\mathrm{s} \, \Gamma (1+z)^{-1}
\label{eq:cool}
\end{equation}
Equations (\ref{eq:1}) and (\ref{eq:cool}) together yield the comoving values for $\gamma_{c,1}$ and $B_1$. 

%
%which gives a value of the emission radius, $R$. %These time scales should be compared to the shortest spikes in the light curve for a consistency check of the assumptions above.

Next, the measured ratio of 
\begin{equation}
{\frac{\nuc}{\num}} = \left( \frac{\gc}{\gm} \right)^2
\end{equation}
which gives a value of ${\gm}_{1}$.

Turning over to interval 2, as mentioned above, we will assume that the cooling times are the same in both intervals (the timebins are of similar widths), which is equivalent of the same emission radius, $R$. Consequently,
\begin{equation}
\frac{t_{c,1}}{t_{c,2}} = \frac{B_2^2 \gamma_{c, 2} }{B_1^2 \gamma_{c, 1}} = 1
\label{eq:4}
\end{equation}
This can be combined with 
\begin{equation}
\frac{\gamma_{\rm{c} 1}}{\gamma_{\rm{c} 2}}  = \left(\frac{{\nuc}_1}{{\nuc}_2} \frac{B_2}{B_1}\right)^{1/2}
\end{equation}
from eq. (\ref{eq:1}), which gives with eq. (\ref{eq:4})
\begin{equation}
\frac{B_2}{B_1} = \left(\frac{{\nuc}_1}{{\nuc}_2}\right)^{1/3}
\label{eq:6}
\end{equation}
giving $B_2$ and 
\begin{equation}
\frac{{\gc}_2}{{\gc}_1}  = \left(\frac{{\nuc}_2}{{\nuc}_1}\right)^{2/3}
\end{equation}
giving ${\gc}_2$.

From eq. (\ref{eq:1}) we similarly have that
\begin{equation}
{\frac{{\nuc}_2}{{\num}_2}} = \left( \frac{{\gc}_2}{{\gm}_2} \right)^2
\end{equation}
which gives a value of $\gamma_{\rm m 2}$. 

For the energy flux at the spectral peak ($\nuc$) with ${\F}_1 \equiv F_\nu({\nuc}_1)$ we have
\begin{equation}
{\frac{{\F}_1}{{\F}_2}} = \frac{{\Pn}_1 N_{e1}}{{\Pn}_2  N_{e2}}
\end{equation}
which, since $P_{\nu, c} \propto B$, combined with eq. (\ref{eq:6}), gives
\begin{equation}
{\frac{N_{e1}}{N_{e2}}} = \frac{{\F}_1}{{\F}_2} \left(\frac{{\nuc}_1}{{\nuc}_ 2}\right)^{1/3}
\end{equation}
\noindent
The number of electrons is (approximate to a factor of a few)
\begin{equation}
N_{e1} \sim \frac{4 \pi d_L^2}{(1+z)} \frac{q}{\me c^2 \sigma_{\rm T}} {\frac{{\F}_{1}}{B_1 \Gamma}}
\end{equation}

\noindent
From this we can calculate the  magnetisation, which is defined as the ratio of the Poynting flux and the matter enthalpy flux, and  becomes in the downstream of the shock
\begin{equation}
%\sigma_1 = \frac{B_1^2}{4 \pi \rho c^2} = 
\sigma_{\rm d} (r) = \frac{B'^2}{4 \pi \Gamma n' (r) \, m_{\rm p} c^2}, %=  \frac{B_1^2}{4 \pi A_1  c^2} R^{2}
\label{eq:sigma}
\end{equation}
where $n'(r)$ is the comoving number density of the radiating electrons  at radius $r$. {Both $n'$, and $B'$ are determined in the downstream, where the energy has been dissipated. }% Therefore, $\sigma_{\rm d}$ is also for the downstream value.}  
The $N_{\rm e}$ electrons radiate from volume $\pi (r/\Gamma)^2 \Delta r'$, where $\Delta r' \sim ct'_{\rm c}/3$. Therefore, 
\begin{equation}
n' = \frac{3 N_{\rm e} \Gamma^2 }{\pi r^2 c t'_{c}}    
\end{equation}
and
\begin{equation}
\sigma_{\rm d} = \frac{B'^2 \, t'_{\rm c}\, r^2 }{12 \mpp c \, N_{\rm e} \Gamma^3} = \frac{B'^2 \, t^{\rm obs}_{\rm c}\, r^2 }{12 \mpp c \, N_{\rm e} \Gamma^2 (1+z)}   
\label{eq:sigma14}
\end{equation}

\vskip 3mm
{Finally, we assume that the large scale, ordered magnetic field is oriented predominantly transverse to the shock normal \citep{Medvedev&Loeb1999, Frederiksen2004, Pelletier_etal_2017}. While the transverse field is largely amplified due to the compression across the shock, any parallel component of the field remains unchanged. The latter component will therefore be largely subdominant just behind the shock \citep{Granot2003}.  In such a case, the magnetisation in the upstream and downstream are largely similar to each other, $\sigma_{\rm d} \sim 3 \sigma_{\rm u}$ \citep[e.g., ][]{Lemoine_Pelletier2010}.}

%%%%%%%%
%Note that with $\gamma_{\rm c}$ is independent of $\Gamma$ (eq. 8). Then the product  $B \Gamma$ is a constant (eq. 1) and therefore $N_e$ is independent of the assumed $\Gamma$ (eq. 12). %(We might want to change the assumption of $\gamma_{\rm c}$, though)
%%%%%%%%%%

%#################################################

\section{Onset of particle acceleration and magnetisation} 
\label{app:2}

A detection of the onset of particle acceleration could indicate that the critical value of the magnetisation $\sigma_{\rm c}$ has been reached and therefore this value can be compared to the observed magnetisation, $\sigma_{\rm d}$. %,  which is, however, measured in the downstream and depends on the bulk Lorentz factor, $\Gamma$.  
%For an individual shock, such as a collision between two shells with different Lorentz factors, $\Gamma_1$ and $\Gamma_2$, the comoving number density of the radiating electrons, $n'$, and $B'$ are determined in the downsteam, where the energy has been dissipated. Therefore, the magnetisation in eq. (\ref{eq:sigma}) is also given for the downstream. 
%On the other hand, the critical value of the magnetisation for the onset of Fermi acceleration, 
The critical value  $\sigma_{\rm c}$, is given in the upstream of the shock as $\Gamma_{\rm r}^2 \sigma_{\rm c, u} \chi_{\rm e}^{-1} \sim 1$, where $\Gamma_{\rm r}$ is the relative Lorentz factor across the shock, %between the upstream and the downstream,
and $\chi_{\rm e} \sim 0.1$ is the fraction of shock energy carried by the accelerated electrons \citep{Lemoine_etal_2013}.
%Therefore when comparing it to observations the $\xi$ factor needs to be considered. 
%The upstream magnetisation, from the observations, can be estimated by using the jump condition for relativistic shocks. 
%In order to compare these values we need to take into account the field amplification across the shock. 
%The compression factor $\xi = n_{\rm d}/n_{\rm u} = 4 \Gamma_{\rm r} + 3$, where $n_{\rm d}$ and $n_{\rm u}$ are the number densities of the electrons in the downstream and upstream, respectively \citep{Blandford1976}. 
%, and $\Gamma_{\rm r}$ is the relative Lorentz factors between the upstream and the downstream.% $\Gamma_{\rm r} \sim 1/2(\Gamma_2/\Gamma_1)^{1/2}$, for $\Gamma_2 \geq \Gamma_1$. 
%The downstream {(transverse)} magnetic field is compressed and amplified as $B_{\rm d} / B_{\rm u} =  \xi$ and therefore, $\sigma_{\rm d} = \sigma_{\rm u} \xi$. 
{For a transverse shock the magnetisation in the upstream and downstream are largely similar and, therefore,} the condition for Fermi acceleration becomes  $\sigma_{\rm d} \simleq \sigma_{\rm c, d} = 0.3  \Gamma_{\rm r}^{-2}$. % for large $\Gamma_{\rel r}$. For external shocks $\Gamma_{\rm r} \sim \Gamma$. % and $\xi \sim  4 \Gamma$.

\begin{figure}[ht!]
\includegraphics[width=\columnwidth]{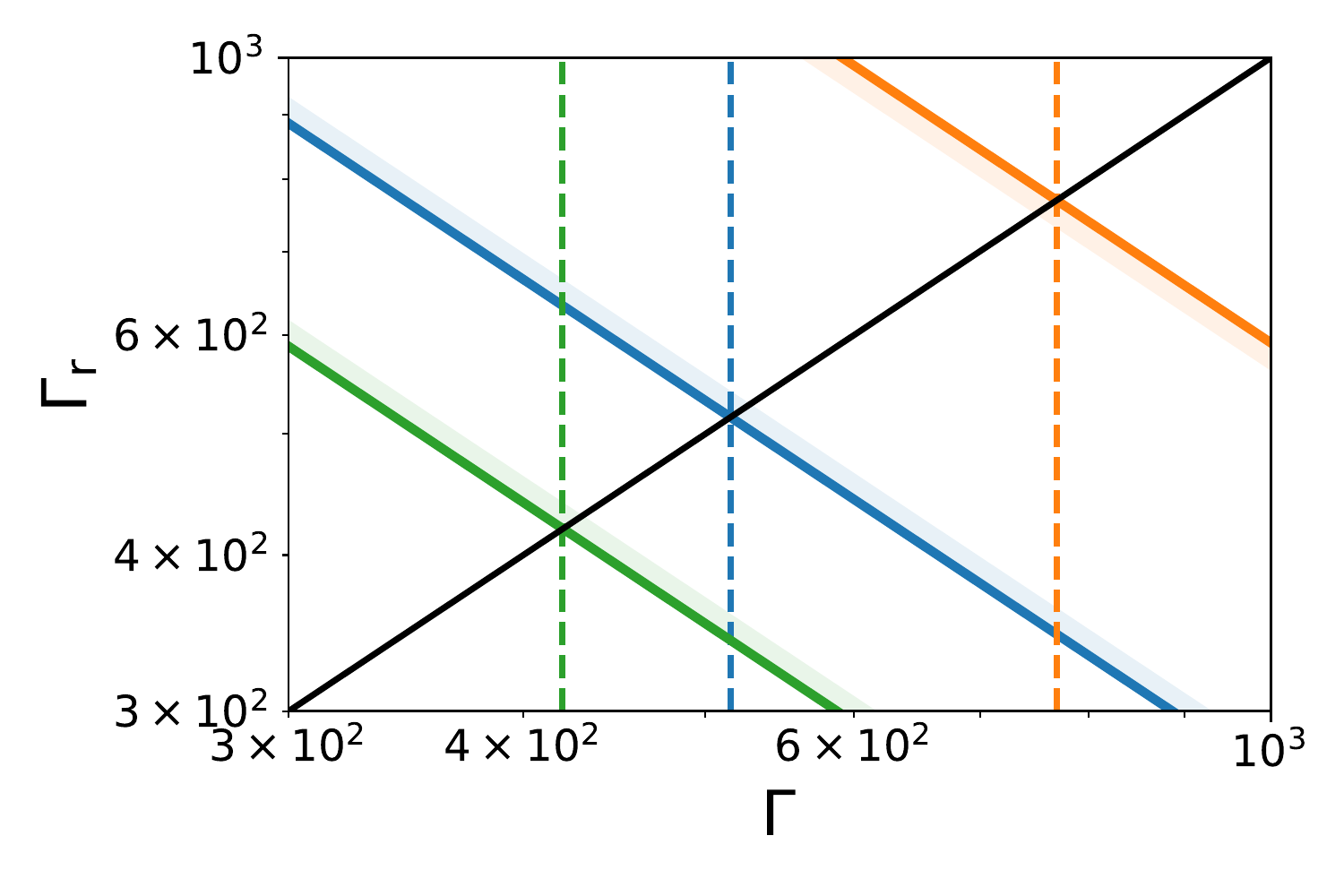}
\caption{{The relative Lorentz factor $\Gamma_{\rm r}$ across the shock versus the bulk Lorentz factor, $\Gamma$.} 
%is the theoretical predicted dependency of the critical magnetisations $\sigma_{u, c}$\cite{Lemoine_Pelletier2010} and 
The colored lines are given by $\Gamma_{\rm r} = 0.55\sigma_{\rm d}^{-1/2}(\Gamma)$, where $\sigma_{\rm d} (\Gamma)$ is the measured magnetisation for different $\Gamma$.
The lines are for the three intervals in GRB160820A  (same colors as in Fig. \ref{fig:1}). The black line is for $\Gamma_{\rm r} = \Gamma$, therefore only parameter values below this line are allowed.  The lines cross each other at $\Gamma =$ 420, 510, and $770$, as marked by the dashed lines. This gives the estimate of the Lorentz factor to be $420 < \Gamma < 770$. A redshift of $z=0.4$ is assumed. %The colored lines are for other assumed redshift values from $z=0.5$ to 2.5 in steps of 0.5.
}
\begin{comment}
\caption{{\bf The upstream magnetisation as a function of bulk Lorentz factor.} The black line is the theoretical predicted dependency of the critical magnetisations $\sigma_{u, c}$\cite{Lemoine_Pelletier2010} and the colored lines are the magnetisation in the three intervals in GRB160820A  (same colors as in Fig. \ref{fig:1}) derived from the downstream $\sigma_{\rm u} = \sigma_{\rm d} (\xi/10)^{-1}$, where compression factor of $\xi=10$ is assumed. The lines cross each other at $\Gamma =$ 715, 880, and $1300$, as marked by the dashed lines. This gives the estimate of the Lorentz factor to be $715 < \Gamma < 1300$. A redshift of $z=0.4$ is assumed. %The colored lines are for other assumed redshift values from $z=0.5$ to 2.5 in steps of 0.5.
}
\end{comment}
\label{fig:sigma}
\end{figure}

 A comparison between the measured magnetisation $\sigma_{\rm d}$ and the critical value $\sigma_{\rm c, d}$ for Interval 2 (the interval with particle acceleration) gives an upper limit of $\Gamma_{\rm r} \leq 0.55  \sigma_{\rm d}^{-1/2}(\Gamma)$. For the other two intervals the requirement give lower limits. 
% For interval 2, which has particle acceleration, this gives an upper limit of $\Gamma_{\rm r} \leq (3  \xi_{\rm e}/ \sigma)^{1/2}$. For the other two intervals the requirement give lower limits. 
Figure \ref{fig:sigma} illustrates this for the three intervals in GRB~160821A. Since $\Gamma_{\rm r}$ by necessity is smaller than $\Gamma$, this analysis gives a lower range of the estimated bulk Lorentz factor of $420  \simleq \Gamma  \simleq 770$ (assuming redshift $z = 0.4$).  

{The range of Lorentz factors is consistent with the values estimated from the general correlation between $E_{\rm iso}$ and $\Gamma$ found from afterglow measurements \citep{Liang2010}: For the estimated value of $E_{\rm iso}\sim 6.9 \times 10^{53}$ erg for GRB160821A (App. \ref{app:115}), the expected range is $300 \simleq \Gamma \simleq 900$.} 
We note that such large values of  $\Gamma$ are also typically found for synchrotron fits in GRBs using other methods to determine its value \citep{Kumar&McMahon2008, Beniamini&Piran2013, Beniamini_etal_2014, Iyyani2016, burgess2020}. 
{Finally, the physical parameter values from the synchrotron fit  assuming the averaged value over the Lorentz factor range, $\Gamma = 595$, are shown in Table \ref{tab:4}.}

We also note that the high-energy slopes of synchrotron spectra in other GRBs are typically softer than what is expected from Fermi acceleration \citep{Goldstein2012, Yu2016, Ravasio2018, Yu2019, burgess2020, Li2021}. This indicates that $\sigma_{c}$ must be low in order for the magnetisation of the GRB to inhibit the particle acceleration. This in its turn again requires that $\Gamma_{\rm r}$ must be large for these bursts \citep{Lemoine_Pelletier2010}, suggesting once more an external shock origin of such emission. 
%For interval 2, the crossing at $\sigma = \sigma_{\rm u, c}$ gives an upper limit on $\Gamma$, while for the other two intervals the crossing give lower limits. 
%Therefore, the estimate of $\Gamma$ in GRB~160821A gives the range 
%$250 < \Gamma < 400$ (assuming redshift $z=1$).
%we find an estimate gives the value , indicated by the red dashed line in Fig. \ref{fig:sigma}. %%% We note that this value is close to the fiducial value we selected. %If instead, $\Gamma$ is determined by an independent method the redshift can instead be constrained. 

\begin{table*}
	\centering
		\begin{tabular}{*{8}{l}}
%\begin{tabular}{|c|c|c|c|c|c|c|c|c|c|}
	% four columns, alignment for each
		\hline
		 Time intervals &
		 $\gamma_{\rm c}$ & $\gamma_{\rm m}$ & $B$ &
		 $R_{\rm dyn}$ &  $N_{e}$  & $\sigma$\\ 
		 & ($10^{5}$) & ($10^{5}$) & (Gauss) & ($10^{17}$ cm) & ($10^{49}$) &  ($10^{-7}$)\\ 
		 \hline
		 \\
		 Interval 1 & $1.06^{+0.08}_{-0.08} $ & $7.5^{+0.5}_{-0.5}$ & $1.31^{+0.05}_{-0.05}$ & $8.5$ & $2.74^{+0.06}_{-0.05}$  & $15.2^{+1.4}_{-1.5} $ \\ 
		 \\		 
		 Interval 2 & $1.41^{+0.06}_{-0.07}$ & $5.2^{+0.8}_{-0.8}$ & $1.14^{+0.03}_{-0.03}$ & "     & $10.2^{+0.1}_{-0.1}$  & $3.0^{+0.2}_{-0.2}$ \\ 
        \\
		 Interval 3 & $0.53^{+0.04}_{-0.05}$ & $5.0^{+0.5}_{-0.5}$ & $1.85^{+0.10}_{-0.07}$ & "     & $2.40^{+0.07}_{-0.06}$   & $34.5^{+3.8}_{-4.7}$  \\
	\hline
	\end{tabular}
	\caption{Derived physical parameters for $\Gamma= 595$ and $z=0.4$. The error intervals are reported with 95\% confidence interval and estimated as Bayesian credible intervals. {The critical value of the magnetisation is $\sigma_c = 8.5 \times 10^{-7}$.}
	}
	\label{tab:4} 
\end{table*}

\end{document}